\documentclass[%
aps,
prl,
twocolumn,
superscriptaddress,
amsmath,amssymb,
nolinenumbers,
]{revtex4-1}
\usepackage[utf8]{inputenc}
\usepackage{graphicx}
\usepackage{dcolumn}
\usepackage{bm}
\usepackage{hyperref}
\usepackage[T1]{fontenc}
\hypersetup{colorlinks=true,
            allcolors=black,
            bookmarksopen=false,
            bookmarksnumbered
           }

\begin{document}

\title{Measurement of the Parity-Violating Asymmetry in the $N\rightarrow \Delta$ Transition at Low $Q^2$}

\author{D.~Adhikari}
\affiliation{Virginia Polytechnic Institute \& State University, Blacksburg, VA 24061, USA}
\author{T. Alshayeb}
\affiliation{Louisiana Tech University, Ruston, LA 71272, USA}
\author{D.~Androi\'c}
\affiliation{University of Zagreb, Zagreb, HR 10002, Croatia }
\author{D.S.~Armstrong}
\affiliation{William and Mary, Williamsburg, VA 23185, USA}
\author{A.~Asaturyan}
\affiliation{A.~I.~Alikhanyan National Science Laboratory (Yerevan Physics Institute), Yerevan 0036, Armenia}
\affiliation{Thomas Jefferson National Accelerator Facility, Newport News, VA 23606, USA}
\author{K.~Bartlett}
\affiliation{William and Mary, Williamsburg, VA 23185, USA}
\author{R.S.~Beminiwattha}
\affiliation{Ohio University, Athens, OH 45701, USA}
\affiliation{Louisiana Tech University, Ruston, LA 71272, USA}
\author{J.~Benesch}
\affiliation{Thomas Jefferson National Accelerator Facility, Newport News, VA 23606, USA}
\author{F.~Benmokhtar}
\affiliation{Duquesne University, Pittsburgh, Pennsylvania 15282, USA}
\author{R.D.~Carlini}
\affiliation{Thomas Jefferson National Accelerator Facility, Newport News, VA 23606, USA}
\affiliation{William and Mary, Williamsburg, VA 23185, USA}
\author{J.C.~Cornejo}
\affiliation{William and Mary, Williamsburg, VA 23185, USA}
\author{S.~Covrig Dusa}
\affiliation{Thomas Jefferson National Accelerator Facility, Newport News, VA 23606, USA}
\author{M.M.~Dalton}
\affiliation{University of Virginia,  Charlottesville, VA 22903, USA}
\affiliation{Thomas Jefferson National Accelerator Facility, Newport News, VA 23606, USA}
\author{C.A.~Davis}
\affiliation{TRIUMF, Vancouver, BC V6T2A3, Canada}
\author{W.~Deconinck}
\affiliation{William and Mary, Williamsburg, VA 23185, USA}
\author{J.A.~Dunne}
\affiliation{Mississippi State University,  Mississippi State, MS 39762, USA}
\author{D.~Dutta}
\affiliation{Mississippi State University,  Mississippi State, MS 39762, USA}
\author{W.S.~Duvall}
\affiliation{Virginia Polytechnic Institute \& State University, Blacksburg, VA 24061, USA}
\author{M.~Elaasar}
\affiliation{Southern University at New Orleans, New Orleans, LA 70126, USA}
\author{W.R.~Falk}
\thanks{Deceased}
\affiliation{University of Manitoba, Winnipeg, MB R3T2N2, Canada}
\author{J.M.~Finn}
\thanks{Deceased}
\affiliation{William and Mary, Williamsburg, VA 23185, USA}
\author{C.~Gal}
\affiliation{University of Virginia,  Charlottesville, VA 22903, USA}
\author{D.~Gaskell}
\affiliation{Thomas Jefferson National Accelerator Facility, Newport News, VA 23606, USA}
\author{M.T.W.~Gericke}
\affiliation{University of Manitoba, Winnipeg, MB R3T2N2, Canada}
\author{J.R.~Hoskins}
\affiliation{William and Mary, Williamsburg, VA 23185, USA}
\author{D.C.~Jones}
\affiliation{University of Virginia,  Charlottesville, VA 22903, USA}
\author{P.M.~King}
\affiliation{Ohio University, Athens, OH 45701, USA}
\author{E.~Korkmaz}
\affiliation{University of Northern British Columbia, Prince George, BC V2N4Z9, Canada}
\author{S.~Kowalski}
\affiliation{Massachusetts Institute of Technology,  Cambridge, MA 02139, USA}
\author{J.~Leacock}
\affiliation{Virginia Polytechnic Institute \& State University, Blacksburg, VA 24061, USA}
\author{J.P.~Leckey}
\affiliation{William and Mary, Williamsburg, VA 23185, USA}
\author{A.R.~Lee}
\affiliation{Virginia Polytechnic Institute \& State University, Blacksburg, VA 24061, USA}
\author{J.H.~Lee}
\affiliation{William and Mary, Williamsburg, VA 23185, USA}
\affiliation{Ohio University, Athens, OH 45701, USA}
\author{L.~Lee}
\affiliation{TRIUMF, Vancouver, BC V6T2A3, Canada}
\affiliation{University of Manitoba, Winnipeg, MB R3T2N2, Canada}
\author{S.~MacEwan}
\affiliation{University of Manitoba, Winnipeg, MB R3T2N2, Canada}
\author{D.~Mack}
\affiliation{Thomas Jefferson National Accelerator Facility, Newport News, VA 23606, USA}
\author{J.A.~Magee}
\affiliation{William and Mary, Williamsburg, VA 23185, USA}
\author{R.~Mahurin}
\affiliation{University of Manitoba, Winnipeg, MB R3T2N2, Canada}
\author{J.~Mammei}
\affiliation{Virginia Polytechnic Institute \& State University, Blacksburg, VA 24061, USA}
\affiliation{University of Manitoba, Winnipeg, MB R3T2N2, Canada}
\author{J.W.~Martin}
\affiliation{University of Winnipeg, Winnipeg, MB R3B2E9, Canada}
\author{M.J.~McHugh}
\affiliation{George Washington University, Washington, DC 20052, USA}
\author{K.E.~Mesick}
\affiliation{George Washington University, Washington, DC 20052, USA}
\affiliation{Rutgers, The State University of New Jersey, Piscataway, NJ 08854, USA}
\author{R.~Michaels}
\affiliation{Thomas Jefferson National Accelerator Facility, Newport News, VA 23606, USA}
\author{A.~Micherdzinska}
\affiliation{George Washington University, Washington, DC 20052, USA}
\author{A.~Mkrtchyan}
\affiliation{A.~I.~Alikhanyan National Science Laboratory (Yerevan Physics Institute), Yerevan 0036, Armenia}
\author{H.~Mkrtchyan}
\affiliation{A.~I.~Alikhanyan National Science Laboratory (Yerevan Physics Institute), Yerevan 0036, Armenia}
\author{L.Z.~Ndukum}
\affiliation{Mississippi State University,  Mississippi State, MS 39762, USA}
\author{H.~Nuhait}
\affiliation{Louisiana Tech University, Ruston, LA 71272, USA}
\author{Nuruzzaman}
\affiliation{Hampton University, Hampton, VA 23668, USA}
\affiliation{Mississippi State University,  Mississippi State, MS 39762, USA}
\author{W.T.H van Oers}
\affiliation{TRIUMF, Vancouver, BC V6T2A3, Canada}
\affiliation{University of Manitoba, Winnipeg, MB R3T2N2, Canada}
\author{S.A.~Page}
\affiliation{University of Manitoba, Winnipeg, MB R3T2N2, Canada}
\author{J.~Pan}
\affiliation{University of Manitoba, Winnipeg, MB R3T2N2, Canada}
\author{K.D.~Paschke}
\affiliation{University of Virginia,  Charlottesville, VA 22903, USA}
\author{S.K.~Phillips}
\affiliation{University of New Hampshire, Durham, NH 03824, USA}
\author{M.L.~Pitt}
\affiliation{Virginia Polytechnic Institute \& State University, Blacksburg, VA 24061, USA}
\author{R.W. Radloff}
\affiliation{Ohio University, Athens, OH 45701, USA}
\author{J.F.~Rajotte}
\affiliation{Massachusetts Institute of Technology,  Cambridge, MA 02139, USA}
\author{W.D.~Ramsay}
\affiliation{TRIUMF, Vancouver, BC V6T2A3, Canada}
\affiliation{University of Manitoba, Winnipeg, MB R3T2N2, Canada}
\author{J.~Roche}
\affiliation{Ohio University, Athens, OH 45701, USA}
\author{B.~Sawatzky}
\affiliation{Thomas Jefferson National Accelerator Facility, Newport News, VA 23606, USA}
\author{N.~Simicevic}
\affiliation{Louisiana Tech University, Ruston, LA 71272, USA}
\author{G.R.~Smith}
\affiliation{Thomas Jefferson National Accelerator Facility, Newport News, VA 23606, USA}
\author{P.~Solvignon}
\thanks{Deceased}
\affiliation{Thomas Jefferson National Accelerator Facility, Newport News, VA 23606, USA}
\author{D.T.~Spayde}
\affiliation{Hendrix College, Conway, AR 72032, USA}
\author{A.~Subedi}
\affiliation{Mississippi State University,  Mississippi State, MS 39762, USA}
\author{W.A.~Tobias}
\affiliation{University of Virginia,  Charlottesville, VA 22903, USA}
\author{V.~Tvaskis}
\affiliation{University of Winnipeg, Winnipeg, MB R3B2E9, Canada}
\affiliation{University of Manitoba, Winnipeg, MB R3T2N2, Canada}
\author{B.~Waidyawansa}
\affiliation{Ohio University, Athens, OH 45701, USA}
\affiliation{Louisiana Tech University, Ruston, LA 71272, USA}
\author{P.~Wang}
\affiliation{University of Manitoba, Winnipeg, MB R3T2N2, Canada}
\author{S.P.~Wells}
\affiliation{Louisiana Tech University, Ruston, LA 71272, USA}
\author{S.A.~Wood}
\affiliation{Thomas Jefferson National Accelerator Facility, Newport News, VA 23606, USA}
\author{P.~Zang}
\affiliation{Syracuse University, Syracuse, New York 13244, USA}
\author{S.~Zhamkochyan}
\affiliation{A.~I.~Alikhanyan National Science Laboratory (Yerevan Physics Institute), Yerevan 0036, Armenia}

\collaboration{Q$_{\text{weak}}$ Collaboration}

\date{April 17 2025}

\begin{abstract}
  
 We report the measurement of the parity-violating asymmetry in the $N\rightarrow \Delta$ transition via the $e^- + p \rightarrow e^- + \Delta ^+$ reaction at two different kinematic points with low four-momentum transfer $Q^2$. Measurements were made with incident electron beam energies of 0.877 and 1.16 GeV, corresponding to $Q^2$ values 
 of 0.0111 and 0.0208 (GeV/c)$^2$, respectively. These measurements put constraints on a low-energy constant in the weak Lagrangian, $d_{\Delta}$, corresponding to a parity-violating electric-dipole transition matrix element. This matrix element has been shown to be large in the strangeness-changing channel, via weak hyperon decays such as $\Sigma ^+ \rightarrow p\gamma$. The measurements reported here constrain $d_{\Delta}$ in the strangeness-conserving channel. The final asymmetries  were $-0.65 \pm 1.00 ({\rm stat.}) \pm 1.02 ({\rm syst.})$ ppm (parts per million) for 0.877 GeV and $-3.59 \pm 0.82 ({\rm stat.}) \pm 1.33 ({\rm syst.})$ ppm for 1.16 GeV. With these results we deduce a  small value for $d_{\Delta}$, consistent with zero, in the strangeness-conserving channel, in contrast to the large value for $d_{\Delta}$ previously reported in the strangeness-changing channel.

\end{abstract}

\maketitle

{\it Introduction/Motivation} --
The $\Delta$(1232) resonance, the first excited state of the proton,
has often been used as a testing ground for 
QCD-inspired models of hadron
structure, as well as underlying QCD symmetries. Many experimental studies
of the excitation of this resonance have been performed using electromagnetic
and strong probes~\cite{Pascalutsa:2006up}. Far fewer studies have been performed with weak probes.
Fewer still have been excitations in the neutral sector of the weak interaction,
i.e., with the exchange of a $Z_0$ boson~\cite{Androic12,G0:2012hkh}. These neutral weak excitations can be
accessed through parity-violating electron scattering experiments from the proton,
where the proton is excited to the $\Delta ^+$ resonance. We report the measurement
of the parity-violating excitation of the $\Delta ^+$ in electron-proton scattering as part of the $Q_{\rm weak}$ experiment \cite{QweakNature}
performed at the Thomas Jefferson National Accelerator Facility in Newport News, Virginia.

The parity-violating (PV) asymmetry in electron proton scattering, in this case for the production of the $\Delta$, is 
\begin{equation}
A_{N\Delta}~=~{{\sigma _+ - \sigma _-} \over {\sigma _+ +\sigma _-}}
\end{equation}
where $\sigma _{+(-)}$
is the cross section for scattering electrons of positive
(negative) helicity (where the electron beam polarization is parallel (anti-parallel) to the beam momentum).
This can be expressed in terms of inelastic response functions as \cite{Zhu012}
\begin{equation}
A_{N\Delta}~=~-{{G_F}\over{\sqrt{2}}} {{Q^2}\over{2\pi\alpha}}{[\Delta_{(1)}^{\pi}
+ \Delta_{(2)}^{\pi} + \Delta_{(3)}^{\pi}]},
\end{equation}
where $\alpha$ is the electromagnetic coupling constant, $G_F$ is the
Fermi constant, and $Q^2$ is the four-momentum transfer.
Here, $\Delta_{(1)}^{\pi} = (1-2\sin ^2 \theta _W)$ is the isovector weak charge
with $\theta _W$ the weak mixing angle,
$\Delta_{(2)}^{\pi}$ contains nonresonant background terms, and $\Delta_{(3)}^{\pi}$
is the isovector, axial-vector nucleon response during its transition to the $\Delta$
resonance.

In addition to the nonresonant contribution 
in $\Delta_{(2)}^{\pi}$,
which was analyzed in detail in \cite{Muk98},
 weak radiative corrections which 
contribute to $\Delta_{(3)}^{\pi}$ 
must be taken into account.
During an investigation of these corrections
to the PV asymmetry in the $N\rightarrow \Delta$ transition,
the authors of Ref. \cite{Zhu012}, using a QCD-inspired model
in a heavy-baryon chiral perturbation theory (HB$\chi$PT) formalism,
uncovered a new type of
radiative correction for inelastic reactions which does not
contribute to elastic scattering.
Although originating from the same Feynman diagram
describing the so-called ``anapole'' contributions (i.e., a
photon coupling to a PV hadronic vertex), one
correction involves a PV
$\gamma N \Delta$ electric-dipole transition, which has no
analog in the elastic channel. As a consequence of Siegert's
theorem, the leading component from the contribution of this transition amplitude
is $Q^2$-independent,
and is proportional to $\omega ~(\omega = E_f -E_i )$
times the PV E1 matrix element, which is characterized
by a low-energy constant $d_{\Delta}$, and can result in a non-vanishing PV
asymmetry at $Q^2 = 0$ \cite{Zhu012}. Thus, a  measurement of the PV
asymmetry in the $N\rightarrow \Delta$ transition at the photon point,
or at very low $Q^2$, 
provides a direct measurement of the low-energy constant $d_{\Delta}$, and therefore provides a constraint on
the weak Lagrangian for this and other reactions involving $d_{\Delta}$.

The quantity $d_{\Delta}$ is related to other interesting physics. 
As mentioned above, $d_{\Delta}$ is given by the PV E1
matrix element, the same transition which drives the asymmetry
parameters in radiative hyperon decays, e.g., $\Sigma ^+ \rightarrow p\gamma$.
A long-standing puzzle in hyperon decay physics has been to understand
the large, negative values obtained for these parameters, which would
vanish in the exact SU(3) limit, a result known as Hara's theorem \cite{Hara64}.
Although typical SU(3) breaking effects are of the order $(m_s - m_u )/(1.0$~GeV) $\approx$ 15\% \cite{Zhu012},
experimentally the asymmetry parameter for $\Sigma ^+ \rightarrow p\gamma$ is
found to be five times larger. There has been renewed interest in understanding this puzzle
in light of recent hyperon decay measurements at BESIII~\cite{Shi25, Shi2023}.
Borasoy and Holstein~\cite{Bor99, Bor99b} proposed a solution
to this puzzle by including high-mass intermediate-state resonances ($J^P = 1/2 ^- )$,
where the weak Lagrangian allows the coupling of both the hyperon and daughter
nucleon to the intermediate-state resonances, driving the asymmetry parameter
to large negative values. This same reaction mechanism was also shown to
simultaneously reproduce the $s-$ and $p-$ wave amplitudes in nonleptonic
hyperon decays, the simultaneous description of which has also been a puzzle in hyperon decay
physics. Thus, if the same underlying dynamics is present in the non-strange sector ($\Delta S = 0$)
as is present in the strangeness-changing sector ($\Delta S = 1$), it would be expected that $d_{\Delta}$ is enhanced relative to its natural scale ($g_{\pi}$ = 3.8$\times 10^{-8}$,
corresponding to the scale of charged-current hadronic PV effects \cite{Des80,Zhu00}).
The authors of \cite{Zhu012} estimate that this enhancement may be as large as a factor
of 100, corresponding to an asymmetry of $\approx$ 4 ppm, an order of magnitude larger than
the asymmetry obtained for the $Q_{\rm weak}$ elastic measurement \cite{QweakNature}. Thus, the measurement
of this quantity provides a window into the underlying dynamics of
the unexpectedly large SU(3) symmetry-breaking effects seen in hyperon decays.

{\it The Experiment} -- The measurements reported here were performed using a custom apparatus \cite{QweakNIM}
built for the $Q_{\rm weak}$ experiment in Jefferson Lab's Hall C. 
The experimental apparatus and its general performance are thoroughly described in \cite{QweakNIM}; here we present only those details relevant to the extraction of the PV asymmetries in the $N \rightarrow \Delta$ transition.
These measurements were carried
out over three separate running periods: two with beam energy of 1.16 GeV (labeled Run 1 and Run 2, respectively) and one dedicated run at 0.877 GeV.
The 1.16 GeV electron beam currents were 165~$\mu$A for Run 1 and 180~$\mu$A for Run 2, and had purely longitudinal polarization of magnitude 89\%. The 0.877 GeV
beam was limited to 100$~\mu$A, and had a significant (44\%) transverse polarization component, as calculated from beam spin precession through the accelerator knowing the beam energy and magnetic fields. The transverse beam polarization component introduced the largest systematic
uncertainties (through the uncertainties in the transverse asymmetries from both
the proton in the hydrogen target and from aluminum in the target end caps) in
extracting the PV asymmetry at 0.877 GeV. For both beam energies, the scattering angle
was 7.9$^{\circ}$ with an acceptance width of $\pm 3^{\circ}$. The azimuthal angle $\phi$ covered 49\% of 2$\pi$, resulting in a solid
angle of 43 msr. The acceptance-averaged four-momenta $Q^2$ values were 0.0208 and 0.0111 
(GeV/c)$^2$ \cite{footnote}, and the acceptance-averaged 
invariant mass $W$ values were  $1.212 \pm 0.001$ and $1.191 \pm 0.001$ GeV$/c^2$ for beam energies 1.16
and 0.877 GeV, respectively.

The polarization of the electron beam was reversed at a rate of 960 Hz pseudorandom sequence of
``helicity quartets'' ($+ - - +$) or ($- + + -$). 
A half-wave plate in the laser
optics of the polarized source \cite{Source1, Source2} was inserted or removed approximately every 8 hours to reverse the beam
polarity with respect to the rapid reversal control signals. The beam current was measured using radio-frequency resonant cavities,
or beam current monitors (BCMs). Five beam position monitors (BPMs) upstream of the target were used to derive the position and angle
of the beam at the target. Energy changes were measured with an additional BPM placed in a dispersive section of the beam line.

The intrinsic beam diameter of $\approx$ 250 $\mu$m was rastered to a uniform area of 4.0 $\times$ 4.0 mm$^2$ at the unpolarized liquid hydrogen (LH$_2$) target \cite{Target1}. The acceptance of the experiment was defined by three Pb collimators, each with eight sculpted openings. A symmetric array of
four luminosity monitors was placed on the upstream face of the defining (middle) collimator \cite{LeacockThesis}.

A toroidal
magnet, QTor, centered 6.5 m downstream from the target center consisted of eight coils arrayed azimuthally
about the beam axis. The magnet provided 0.89 T$\cdot$m at a setting of 8900 A, the current required to allow elastically scattered
electrons from a 1.16 GeV beam through the acceptance of the collimator-spectrometer system. To perform the inelastic measurements
reported here, the magnet current was reduced to accept those inelastic electrons which excited the $\Delta$ resonance.
For the 1.16 GeV beam, that setting was 6700 A, while for the 0.877 GeV beam, the setting
was 4650 A (the elastic scattering magnet setting for the 0.877 GeV beam was 6800 A).

The magnet concentrated the inelastically scattered electrons onto eight radiation-hard synthetic fused-quartz Cherenkov
detectors arrayed symmetrically about the beam axis~\cite{WangThesis}.
Azimuthal symmetry was a crucial aspect of the experiment's design. It allowed us to minimize systematic errors
from helicity-correlated changes in the beam trajectory and contamination from residual transverse asymmetries for the longitudinally
polarized beam, while also allowing us to map out the sinusoidal dependence of the transverse asymmetry for transversely polarized
beam. Two 100$\times$18$\times$1.25 cm thick bars glued together into 2 m long bars
comprised each of the eight detectors. 
Cherenkov light from the bars was
read out by 12.7 cm diameter low-gain photomultiplier tubes (PMTs) through 
quartz light guides on each end of the bar assembly.
The detectors were equipped with 2 cm thick Pb preradiators which amplified the electron signal and suppressed soft backgrounds.

With scattered inelastic electron rates of $\approx 50$ MHz per detector, a current-mode readout was required. The anode current from each PMT
was converted to a voltage using a custom low-noise preamplifier and digitized with an 18 bit, 500 kHz sampling ADC whose outputs were
integrated every millisecond. A separate PMT base was used to read out the detectors in a counting (individual pulse) mode at much lower
beam currents (0.1 - 200 nA) during calibration runs. During these runs, the response of each detector was measured using a system of
drift chambers \cite{Tracking1} and trigger scintillators \cite{Tracking2} positioned in front of two detectors at a time and removed
during the asymmetry measurements. These counting-mode runs were critical for comparison of measured data rates to simulation rates for all
physics processes generated in the LH$_2$ target and the Al endcaps enclosing the LH$_2$, including pion production -- these
processes had a much larger relative contribution to the detected signal in the $\Delta$ region than at the elastic peak \cite{HendThesis}.
As part of these calibrations, we took short runs in which the magnet current was changed systematically in steps of 200 A from the elastic
peak, through the $\Delta$ peak, and well into the higher resonance region. These ``field scans'' allowed us to measure the shape of the distribution
of particle rates as a function of excitation energy. These were compared to the simulated version of these rates. Similar field scans were performed in
high beam-current mode 
using current-mode readout to map out the 
dependence of the detector signal on magnetic field.
These were also compared to the simulated version of these scans. Together, these field scans were critical in allowing us to
estimate the signal fractions of background processes under the $\Delta$ peak. 

{\it Data Analysis} -- The experimental raw asymmetry $A_{\rm raw}$ was calculated over each helicity quartet from the integrated PMT signal normalized to the beam
charge ($Y_{\pm}$) as $A_{\rm raw} = (Y_+ - Y_- )/(Y_+ + Y_-)$ and averaged over all detectors. 
The $A_{\rm raw}$ values (listed in Table 1 for all three running periods) were corrected for false asymmetries associated with helicity-correlated beam properties to form
the measured asymmetry $A_{\rm meas}$:
\begin{equation}
\footnotesize
A_{\rm meas} = A_{\rm raw} + A_{\rm BCM} + A_{\rm beam} + A_{\rm BB} + A_L + A_T + A_{\rm bias} - A_{\rm blind}.
\end{equation}

$A_{\rm BCM}$ is the false asymmetry induced by helicity-correlated beam 
current differences, and is zero by definition.
This is because this effect is included in the data analyzer in the formation of $A_{\rm raw}$. We include this term here to ensure that the uncertainty on this correction due to the choice of BCM used to normalize the detector signals to the beam charge
is propagated through to the uncertainty in the measured asymmetry. This contribution, along with many others that contribute to all PV electron scattering experiments at Jefferson Lab, was studied in detail in \cite{Adderley2023}. $A_{\rm beam}$ is the false asymmetry induced due to helicity-correlated beam position, angle,
and energy 
\begin{equation}
A_{\rm beam} = \sum_{i=1}^5 \left( {{\partial A_{\rm det}}\over{\partial \chi _i}} \right )\Delta \chi_i.
\end{equation}
where the $\Delta \chi _i$  are the helicity-correlated changes in beam trajectory or energy 
over the helicity quartet, and the slopes $\partial A$/$\partial \chi _i$  
were
determined using linear regression 
applied to natural motion of the beam, as well as from deliberate periodic modulation
of beam properties~\cite{QweakNature}. 
$A_{\rm BB}$ is a correction for the false asymmetry induced byhelicity-correlated background scattered from the beamline. It was estimated using the correlation $\partial A_{\rm det}/\partial A_{\rm lumi}$ between the main detectors and luminosity monitors placed just downstream of the target but out of the acceptance of the spectrometer. The correlation was multiplied by the asymmetry measured in the luminosity monitors for each running period $\Delta A_{\rm lumi}$ 
\cite{QweakNature},
\begin{equation}
A_{\rm BB} = \left( {{\partial A_{\rm det}}\over{\partial A_{\rm lumi}}}\right)  \Delta A_{\rm lumi} .
\end{equation}
$A_{\rm BB}$ depended on beam conditions and was the largest contributor to the total uncertainty on the PV $N \rightarrow \Delta$ asymmetries for both of the 1.16 GeV data sets.
$A_L$ is a correction that takes into account the small nonlinearity in the detector PMT response. $A_T$ accounts for the transverse
component in the longitudinally polarized beam \cite{BuddhiniThesis}.
$A_{\rm bias}$ is a correction that arose from the polarized electrons undergoing primarily Mott scattering from the lead nuclei in the lead preradiators
positioned in front of the quartz detectors. This correction has been documented at length in \cite{QweakNature}.
Finally, $A_{\rm blind}$ is a constant offset (randomly chosen in a range between $\pm$ 50\% of the expected standard model value of the $ep$ elastic asymmetry)
which was added into the data stream and whose value was not known by the collaboration during the data analysis. 
After the data analysis was completed, $A_{\rm blind}$ was subtracted from
the measured asymmetry. All of these false asymmetries are listed in Table 1 for each of the running periods studied here.

The fully-corrected physics asymmetry $A_{N\rightarrow \Delta}$ was obtained using 
the following equation which accounts for electromagnetic radiative corrections,
kinematics normalization, polarization, and backgrounds:
\begin{equation}
A_{N\rightarrow \Delta} = R_{\rm tot} {{A_{\rm meas}/P - \sum_{i=1,3,4,5} f_i A_i}\over{1 - \sum_{i=1}^{5} f_i}}.
\end{equation}
Here $R_{\rm tot} = R_{\rm RC} R_{\rm Det} R_{\rm Acc} R_{Q^2}$, where $R_{\rm RC}$ is a radiative correction deduced from simulations with and without
bremsstrahlung using methods described in Refs. \cite{HAPPEX1,HAPPEX2}, $R_{\rm Det}$ accounts for the measured light variation and
non-uniform $Q^2$ distribution across the detector bars, $R_{\rm Acc}$ is an effective kinematics correction \cite{HAPPEX2} which
corrects the asymmetry from $\langle A(Q^2 ) \rangle$ to $A(\langle Q^2 \rangle )$, and $R_{Q^2}$ represents the precision in
calibrating the central $Q^2$ value. 
$P$ is the longitudinal beam polarization 
determined with Moller~\cite{polarimeter} and Compton~\cite{Compton} polarimeters, which were cross checked against each other~\cite{Magee:2016xqx}.

For each of the backgrounds $b_i = f_i A_i$, $f_i$ is the 
fraction of the total
signal due to background $i$ included in the $N\rightarrow \Delta$ signal, and $A_i$ is the asymmetry for that process. All of these 
background fractions and asymmetries are listed in Table 1 for the three different running periods.

The largest background 
comes from the radiative tail of the elastic peak ($b_4$) which lies underneath the $N\rightarrow \Delta$ peak.
The 
fraction $f_4$ was nearly 75\% for 1.16 GeV, and
nearly 80\% for 0.877 GeV, and was determined by extensive simulations of the magnetic field scans discussed earlier. Fortunately, the elastic asymmetry $A_4$ was measured to high precision ($\approx  4$\%
relative error) at 1.16 GeV in the $Q_{\rm weak}$ experiment. Since the PV asymmetries in both elastic and inelastic scattering have a leading $Q^2$ dependence
which is linear, and the $Q^2$ values are quite close for elastic and inelastic scattering at the $\Delta$ peak, a simple linear scaling in $Q^2$ from
the elastic peak to the $\Delta$ peak (including a small correction to account for bremsstrahlung in the target) allowed us to determine $A_4$.
Another correction comes from the aluminum windows of the
target cell ($b_1$). The cell-window asymmetry $A_1$ was measured in dedicated runs with dummy targets at the magnetic field for the $\Delta$ peak for each beam energy setting,
while the fraction
$f_1$ was determined through simulation. A fraction
($f_2$ associated with $A_{\rm BB}$, discussed above) 
is that due to scattering sources in the beam line. This fraction
was studied in detail for both elastic and inelastic signals \cite{MartyTechNote}.
An extensive study of the background of neutral particles seen by the Cherenkov
detectors \cite{MartyTechNote} determined their 
fraction 
$f_3$ and asymmetry $A_3$. A final correction ($b_5$) was made to account for the $\pi ^-$'s accepted through the spectrometer and into the detectors
in the $\Delta$ region from the LH$_2$ (from two-pion production off the protons) and from the Al endcaps (from single $\pi ^-$ production off the
neutrons in the Al nuclei). 
The fraction $f_5$ was determined
through simulation.
 There are no theoretical estimates found in the literature for $A_5$, the PV asymmetry in pion production
from the proton where the pions are detected, 
so we assign a value corresponding to the leading term
for all PV electron-scattering asymmetries for processes with single 
weak boson exchange $A_0 = -G_F Q^2 /(\sqrt{2} 2\pi \alpha)$, and conservatively assign
an uncertainty of 100\% of this value for each of the $Q^2$ values
studied here.
All background fractions
$f_i$ and 
asymmetries $A_i$ are listed in Table 1 for each running period. The majority of the analysis of the 0.877 GeV data was
performed in \cite{AnnaLeeThesis}. An early analysis of the 1.16 GeV data set
was carried out in \cite{LeacockThesis}, and a later analysis which included
all of the final corrections to the 1.16 GeV asymmetries was performed in \cite{AlShayebThesis}.

\begin{table}[h]
\footnotesize
\begin{center}
\begin{tabular}{llll}
\multicolumn{1}{c}{Quantity}&
\multicolumn{1}{c}{0.877 GeV}&
\multicolumn{1}{c}{1.16 GeV Run 1}&
\multicolumn{1}{c}{1.16 GeV Run 2}\\
\hline\hline
$A_{\rm raw}$ & $-0.076 \pm 0.075$  & $-1.36 \pm 0.22$  & $-0.685 \pm 0.17$  \\
\hline
$A_{\rm BCM}$ & 0 $\pm$ 0.010  & 0 $\pm$ 0.04  & 0 $\pm$ 0.03  \\
$A_{\rm beam}$ & $-0.018 \pm 0.010$  & 0.04 $\pm$ 0.04  & $-0.052  \pm 0.052$ \\
$A_{\rm BB}$ & 0.028 $\pm$ 0.014  & 0.52 $\pm$ 0.24  & 0.093 $\pm$ 0.194  \\
$A_{L}$ & 0.0001 $\pm$ 0.0004  & 0.002 $\pm$ 0.001  & 0.0011 $\pm$ 0.0009  \\
$A_{T}$ & $-0.036 \pm 0.047$  &  0 $\pm$ 0.032  & 0 $\pm$ 0.012  \\
$A_{\rm bias}$ & 0.0022 $\pm$ 0.0016  &  0.0035 $\pm$ 0.0024  & 0.0035 $\pm$ 0.0024 \\
$A_{\rm blind}$ &  0.00669 $\pm$ 0  &$-0.0253 \pm 0$  & 0.00669 $\pm$ 0  \\
\hline
$A_{\rm meas}$ &  $-0.105 \pm 0.091$  & $-0.770 \pm 0.33$   & $-0.645 \pm 0.26$  \\
\hline
$R_{\rm det}$ & 0.9857 $\pm$ 0.0022 &  0.9811 $\pm$ 0.0022 & 0.9811 $\pm$ 0.0022 \\
$R_{\rm RC}$ & 1.01 $\pm$ 0.005 & 1.01 $\pm$ 0.005 & 1.01 $\pm$ 0.005 \\
$R_{\rm Acc}$ & 1 $\pm$ 0.01 & 1 $\pm$ 0.01 & 1 $\pm$ 0.01 \\
$R_{Q^2}$ & 1 $\pm$ 0.0045 & 1 $\pm$ 0.0045 & 1 $\pm$ 0.0045 \\
\hline
$R_{\rm tot}$ & 0.9956 $\pm$ 0.050 & 0.9909 $\pm$ 0.012 & 0.9909 $\pm$ 0.012 \\
\hline
$P$ & 0.788 $\pm$ 0.016 & 0.8585 $\pm$ 0.010 & 0.886 $\pm$ 0.006 \\
$f_1$ & 0.069 $\pm$ 0.0034  & 0.0358 $\pm$ 0.0018 & 0.0358 $\pm$ 0.0018 \\
$A_1$ & 0.36 $\pm$ 0.89 & 1.61 $\pm$ 1.15  & 1.61 $\pm$ 1.15 \\
$f_3$ & 0.0018 $\pm$ 0.0072  & 0.024 $\pm$ 0.020 & 0.024 $\pm$ 0.020 \\
$A_3$ & $-0.17 \pm 0.10$  & $-0.31 \pm 0.12$  & $-0.31 \pm 0.12$  \\
$f_4$ & 0.790 $\pm$ 0.046  & 0.7242 $\pm$ 0.042 & 0.7242 $\pm$ 0.042 \\
$A_4$ & $-0.096 \pm 0.015$  & $-0.174 \pm 0.016$  & $-0.174 \pm 0.016$\\
$f_5$ & 0.0097 $\pm$ 0.0048 & 0.0094 $\pm$ 0.0047 & 0.0094 $\pm$ 0.0047 \\
$A_5$ & $-2.07 \pm 2.07$ & $-3.60 \pm 3.60$  & $-3.60 \pm 3.60$  \\
$f_2$ & 0.0343 $\pm$ 0.0040 & 0.020 $\pm$ 0.008 & 0.020 $\pm$ 0.008 \\
\hline
$A_{N\rightarrow \Delta}$ & $-0.65 \pm 1.43$  & $-4.18 \pm 2.31$  & $-3.28 \pm 2.16$  \\
\hline\hline
\end{tabular}
\end{center}
\caption{
Measured and false asymmetries, background 
fractions and asymmetries, radiative corrections, and all parameters required to
calculate the $N \rightarrow \Delta$ asymmetries in the different kinematics and running periods. All asymmetries have units of parts per million (ppm), while the polarization ($P$), radiative correction terms ($R$'s), and background 
fractions  ($f_i$'s) are absolute values relative to 1.00.}
\end{table}

{\it Results and Summary} -- The final asymmetries 
were obtained using Eqs. 3 and 6 and the results in 
Table 1.  Quoting statistical and systematic errors separately, these are
for 0.877 GeV, $A_{N\rightarrow \Delta}$ = $-0.65 \pm 1.00 ({\rm stat.}) \pm 1.02 ({\rm sys.})$ ppm, for 1.16 GeV (Run 1) $A_{N\rightarrow \Delta}$ = $-4.18 \pm 1.36 ({\rm stat.}) \pm 1.86 ({\rm sys.})$ ppm, and for 1.16 GeV (Run 2) $A_{N\rightarrow \Delta} = 
-3.28 \pm 1.02 ({\rm stat.}) \pm 1.91 ({\rm sys.})$ ppm. Combining the 1.16 GeV results for Runs 1 and 2, taking into account the correlations among the systematic errors for those two running periods gives, $A_{N\rightarrow \Delta}$ = 
$-3.59 \pm 0.82 ({\rm stat.}) \pm 1.33 ({\rm syst.})$ ppm.

Plotting these two values of the PV $N\rightarrow \Delta$ asymmetry as a function of $Q^2$, 
along with predictions of this asymmetry at low $Q^2$ \cite{Zhu012} for different values of $d_{\Delta}$ (see Fig. 1)
gives us a good indication of the importance of this particular radiative correction to the $N\rightarrow \Delta$ transition. 
Based on the position of the two data points at their respective $Q^2$ values  relative to the curves of \cite{Zhu012} for $d_{\Delta}$ = 0 and 25$ g_{\pi}$, we can determine the measured values of $d_{\Delta}$ for the two measured asymmetries separately. We find $d_{\Delta} = (-20 \pm 25 ({\rm stat.}) \pm 26({\rm syst.}) \pm 3 ({\rm theory}))g_{\pi}$ for 0.877 GeV and $d_{\Delta}$ = ($21 \pm 20 ({\rm stat.}) \pm 33 ({\rm syst.}) \pm 3 ({\rm theory}))g_{\pi}$ for 1.16 GeV, where we have added a small theory error to account for the fact that the calculations of \cite{Zhu012} were performed at a beam energy of 0.424 GeV and taking into account the predicted slowly-varying energy dependence \cite{MusolfPC}. Both of the measured values of $d_{\Delta}$ reported here are consistent with $d_{\Delta} = 0$ within errors. We note that the G0 Collaboration has published a value of $d_{\Delta} = (8.1 \pm 23.7 ({\rm stat.}) \pm  8.3 ({\rm syst.}) \pm 0.7 ({\rm theory}))g_{\pi}$ off the neutron via the $\gamma + d \rightarrow \Delta ^0 + p$ reaction \cite{Androic12} at $Q^2$ = 0.0032 (GeV/c)$^2$, which is also consistent with zero. Because the G0 result is from a different reaction 
than the data reported here, we choose not to include the G0 point 
in Fig. 1.

\begin{figure}[htbp]
\centerline{
\scalebox{0.5}[0.525]{\includegraphics{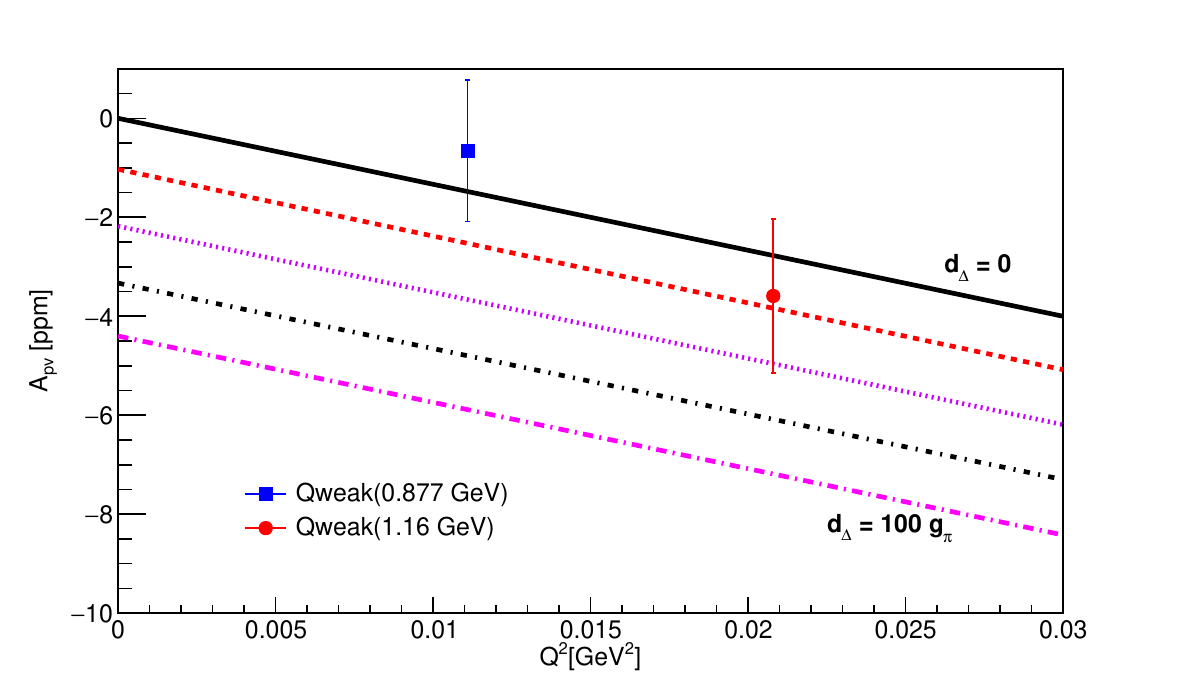}}}
\caption{Plot of the $N \rightarrow \Delta$ asymmetry measurements as a function of $Q^2$ for the measurements reported here, along with the calculations of this asymmetry at low $Q^2$ from \cite{Zhu012} for different values of $d_{\Delta}$ 
ranging from 0 (solid black) to 100 $g_{\pi}$ (dot-dash magenta) in steps of 25$g_{\pi}$.}
\label{fig:AdDelta}
\end{figure}

Combining the statistical, systematic, and theoretical errors from our two measurements of $d_{\Delta}$, we find $d_{\Delta} = (-20 \pm  36)g_{\pi}$ for 0.877 GeV and $d_{\Delta} = (21 \pm 39)g_{\pi}$ for 1.16 GeV. Taking the weighted average of these two independent measurements yields a final value of $d_{\Delta}$ = $(-1.5 \pm 26)g_{\pi}$, which is again 
consistent with $d_{\Delta}$ = 0. 

The PV E1 matrix element characterized by the low-energy constant $d_{\Delta}$ in the weak Lagrangian was proposed to be large in the $\Delta S$ = 1 (strangeness-changing) sector of the weak interaction as evidenced by the large asymmetry parameters seen in weak hyperon decays such as $\Sigma ^+ \rightarrow p\gamma$ which are driven by this matrix element, in contradiction to what standard SU(3) symmetry breaking predicts, yet is found to be small and even consistent with zero in  the $\Delta S$ = 0 (strangeness-conserving) sector as seen in the PV asymmetries in the $N \rightarrow \Delta$ transition reported here. The dynamics included in a QCD-based model \cite{Zhu012} which predict large values of $d_{\Delta}$ and drive the weak hyperon-decay asymmetry parameters to large negative values in better agreement with experiment in the $\Delta S$ = 1 channel do not seem to be present in the PV asymmetries in the $N\rightarrow \Delta$ transition in the $\Delta S$ = 0 channel, which is also driven by the $d_{\Delta}$ matrix element. Thus the QCD-based model which is successful in the $\Delta S$ = 1 channel does not apply in the $\Delta S$ = 0 channel and suggests that different dynamics must be considered for this latter channel.

This work was supported by DOE Contract No. DEAC05-06OR23177, under
which Jefferson Science Associates, LLC operates Thomas Jefferson
National Accelerator Facility. Construction and operating funding for
the experiment was provided through the U.S. Department of Energy
(DOE), the Natural Sciences and Engineering Research Council of Canada
(NSERC), the Canada Foundation for Innovation (CFI), and the
National Science Foundation (NSF) with university matching
contributions from William \& Mary, Virginia Tech, George Washington
U., and Louisiana Tech U. We thank the staff of
Jefferson Lab, in particular the accelerator operations staff, the
target and cryogenic groups, the radiation control staff, as well as
the Hall C technical staff for their help and support. We are grateful
for the contributions of our undergraduate students. We thank TRIUMF
for its contributions to the development of the spectrometer and
integrating electronics, and Bates  Research and Engineering Center for its contributions to the
spectrometer and Compton polarimeter. We are indebted to 
M.J. Ramsey-Musolf
for many useful discussions and insight.


\begin{thebibliography}{}
\bibitem{footnote} The effect of the uncertainty in $Q^2$ is included in the asymmetry via the factor $R_{Q^2}$.
\end{thebibliography}
\end{document}